\documentclass[twocolumn]{aastex631}
\usepackage{amsmath}

\usepackage[T1]{fontenc}

\shorttitle{Molecular gas mass within 500~pc}
\shortauthors{Fujita et al.}

\begin{document}

\title{The relationships between AGN power and molecular gas mass within 500~pc of the center of elliptical galaxies}
\author[0000-0003-0058-9719]{Yutaka Fujita}
\affiliation{Department of Physics, Graduate School of Science
Tokyo Metropolitan University\\
1-1 Minami-Osawa, Hachioji-shi, Tokyo 192-0397, Japan}

\author{Takuma Izumi}
\affiliation{National Astronomical Observatory of Japan, 2-21-1 Osawa, Mitaka, Tokyo 181-8588}  \affiliation{The Graduate
University for Advanced Studies, SOKENDAI, Osawa 2-21-1, Mitaka, Tokyo
181-8588, Japan}

\author{Hiroshi Nagai} \affiliation{National Astronomical Observatory of
Japan, 2-21-1 Osawa, Mitaka, Tokyo 181-8588} \affiliation{The Graduate
University for Advanced Studies, SOKENDAI, Osawa 2-21-1, Mitaka, Tokyo
181-8588, Japan}

\author{Nozomu Kawakatu}
\affiliation{National Institute of Technology, Kure College, 2-2-11, Agaminami, Kure, Hiroshima, 737-8506, Japan}

\author{Norita Kawanaka}
\affiliation{Department of Physics, Graduate School of Science
Tokyo Metropolitan University\\
1-1 Minami-Osawa, Hachioji-shi, Tokyo 192-0397, Japan}
\affiliation{National Astronomical Observatory of
Japan, 2-21-1 Osawa, Mitaka, Tokyo 181-8588} 

\begin{abstract}

The physical quantity that directly controls the feedback of active galactic nuclei (AGNs) in elliptical galaxies remains to be determined.
The discovery of molecular gas around the AGNs suggests that the gas is
fueling the AGNs. Therefore, we analyze Atacama Large
Millimeter/submillimeter Array (ALMA) data for the CO line (J=1--0, 2--1, 3--2)
emission and estimate the mass of molecular gas within 500~pc of the center of 12 non-central elliptical galaxies (NCEGs) and 10 of the brightest cluster galaxies (BCGs). 
We find that the mass ($M_{\rm mol} \sim 10^5$--$10^9\rm\: M_\sun$) is correlated with the jet power of their AGNs, which is represented by $P_{\rm cav}\approx 4.1\times 10^{42}(M_{\rm mol}/10^7\: M_\sun)^{1.3}\rm\: erg\: s^{-1}$,
although NCEGs alone do not show the correlation.
We also find that $M_{\rm mol}$ is correlated with the AGN continuum luminosities at $\sim 1.4$~GHz ($L_{\rm 1.4}$) and $\sim 100$--300~GHz ($L_{\rm con}$). Since $P_{\rm cav}$ reflects galactic-scale, long-term AGN activity, while the continuum luminosities reflect local ($\lesssim 500$~pc), short-term AGN activity, our results suggest that AGN activity depends on the amount of gas, regardless of its time scale.  
On the other hand, we cannot find a clear correlation between the mass of the black holes in the AGNs ($M_{\rm BH}$) and $P_{\rm cav}$.  
This suggests that $M_{\rm mol}$, rather than $M_{\rm BH}$, is the main factor controlling AGN activity. We confirm that the origin of the continuum emission from the AGNs at $\sim 1.4$--300~GHz is mostly synchrotron radiation.

\end{abstract}


\section{Introduction} 
\label{sec:intro}

Supermassive black holes are ubiquitous at the centers of galaxies \citep{2013ARA&A..51..511K}, and they produce an enormous amount of energy as active galactic nuclei (AGNs).  The AGNs in massive elliptical galaxies are in "radio feedback" mode and often show jet activity \citep{2014ARA&A..52..589H}.
Since elliptical galaxies are filled with hot gas \citep{1989ARA&A..27...87F}, it may be natural to assume that the hot gas is fed into the black holes in the form of the Bondi accretion \citep{1952MNRAS.112..195B}. \citet{2006MNRAS.372...21A} showed that there is a strong correlation between the AGN jet power ($P_{\rm cav}$), which can be inferred from the the size of the X-ray cavities in the host galaxies, and the Bondi accretion rate. However, later studies have disproved such a strong correlation. \citep{2013MNRAS.432..530R}.

The discovery of a large amount of molecular gas ($\gtrsim 10^8\rm\: M_\sun$) in elliptical galaxies at the center of nearby clusters \citep{2001MNRAS.328..762E,2003A&A...412..657S,2014ApJ...792...94D,2014ApJ...785...44M,2016ApJ...832..148V,2016MNRAS.458.3134R,2017ApJ...848..101V,2017MNRAS.472.4024R,2019A&A...631A..22O, 2019MNRAS.490.3025R,2020ApJ...894...72S,2020MNRAS.496..364R,2021MNRAS.503.5179N} suggests that this cold gas is feeding the AGNs. 
However, the cold gas is often widely distributed on a scale of $\gtrsim 10$~kpc, and it is unlikely that all of the gas contributes to AGN activity because some of it is consumed in star formation \citep{2022ApJ...924...24F}.  
As direct evidence for AGN feeding, many absorption lines have been observed in some AGNs, indicating the presence of dense gas in the vicinity of the AGNs \citep{2014ApJ...792...94D,2016Natur.534..218T,2019MNRAS.485..229R,2023MNRAS.518..878R}. 
However, it is difficult to estimate the mass of the circumnuclear gas from the absorption.
Thus, it is important to focus on the emission from the cold gas in the vicinity (say $\lesssim 500$~pc) of the AGNs, which can directly influence the gas accretion onto them. 

\begin{figure}[t]
\plotone{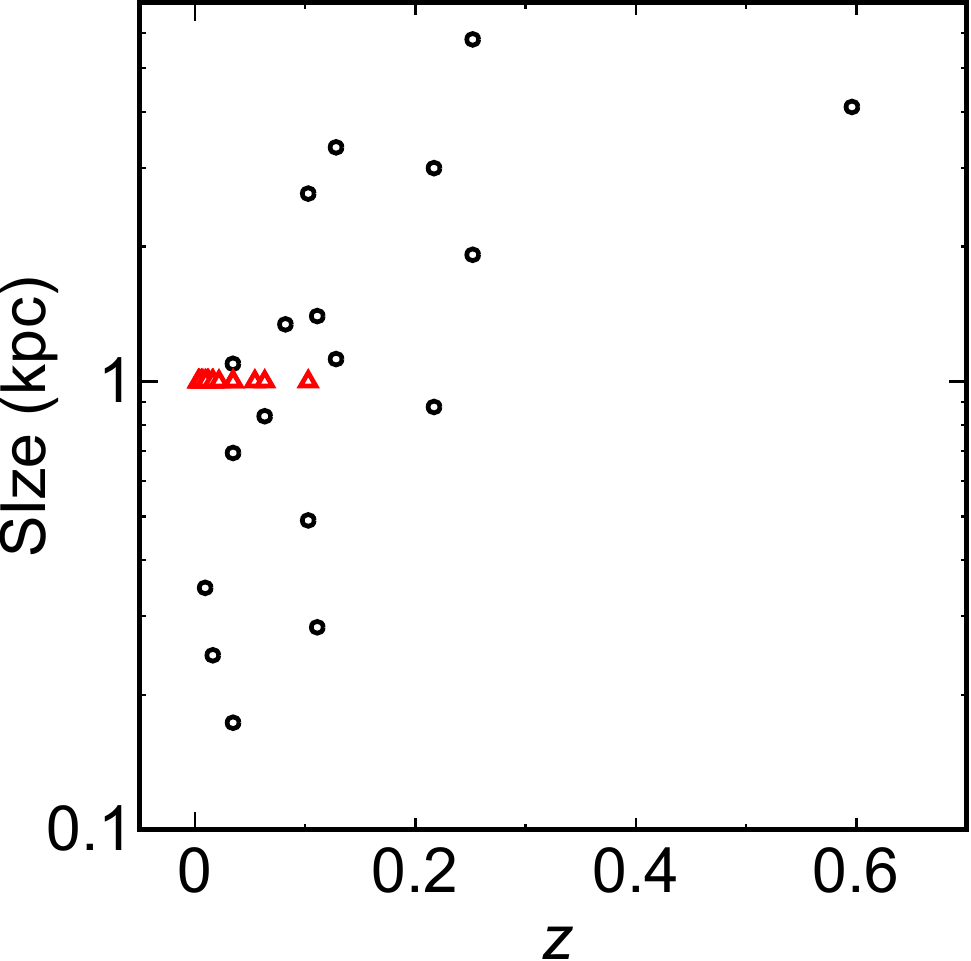} 
\caption{Red open triangles show the redshifts and the diameter of the focused region (1~kpc) for our 22 sample galaxies. Black open circles are the redshifts and geometric mean beam sizes for the galaxies studied by \citet{2019MNRAS.490.3025R}. Note that some of these galaxies were observed with different beam sizes and have multiple black open circles for the given redshift. \label{fig:rus19a}}
\end{figure}

We note that \citet{2019MNRAS.490.3025R} compared the mass of the circumnuclear molecular gas obtained with the Atacama Large Millimeter/submillimeter Array (ALMA) with the AGN jet power $P_{\rm cav}$ and showed that there is a correlation (their Figure 7, see also our Figure~\ref{fig:Mmol}(a)). However, they measured the mass in a single ALMA-synthesized beam centered on the AGNs. Since the physical scale of a given beam is larger for more distant objects (see Section~\ref{sec:sample} and Figure~\ref{fig:rus19a}), the mass in that beam is likely to be larger, while distant AGNs are observationally biased to be brighter and more powerful. This could lead to an artifact relationship. 

For this reason, we examined the relationship between AGN power and molecular gas mass within a fixed physical radius ($M_{\rm mol}$), i.e. within 500~pc of the black holes, for 9 brightest cluster galaxies (BCGs) at the center of clusters \citep[Paper I;][]{2023arXiv230316927F}. 
The 9 BCGs are part of the samples of \citet{2019A&A...631A..22O} and \citet{2019MNRAS.490.3025R}.
Since the physical scale of the region is fixed, the above bias should be less likely to occur.  

We note that this scale (500~pc) is comparable to the circumnuclear disk around the black hole \citep{2019ApJ...883..193N,2022ApJ...924...24F}. In Paper~I we found that $M_{\rm mol}$ is correlated with the AGN jet power ($P_{\rm cav}$). Since the timescale of cavity formation is $\sim 10^7$~yr \citep{2004ApJ...607..800B,2006MNRAS.372...21A,2006ApJ...652...216R}, $P_{\rm cav}$ is the power averaged over this timescale.  On the other hand, we found that the continuum luminosities of AGNs at 1.4~GHz ($L_{1.4}$) and $\sim 100$--300~GHz ($L_{\rm con}$) are not correlated with $M_{\rm mol}$ (Paper~I). The continuum emission is observed as a point source. Considering that the typical beam size is $\lesssim 500$~pc and the corresponding light travel time is $\lesssim 1600$~yr, the emissions reflect the recent activity of the AGNs. Thus, the lack of correlation between the continuum luminosities and $M_{\rm mol}$ may indicate that AGN activity varies on a short time scale. Indeed, intermittent jet activity and jet-cloud interactions have been observed for NGC~1275, the BCG of the Perseus cluster. This object shows radio luminosity variations on a timescale of decades \citep[e.g.][]{2010PASJ...62L..11N,2017MNRAS.465L..94F,2017ApJ...849...52N,2021ApJ...920L..24K}.

In this study, we extend the sample of galaxies studied in Paper~I to lower masses by including 12 nearby non-central elliptical galaxies (NCEGs) that are not at the center of massive clusters \footnote{There is no strict distinction between NCEGs and BCGs. Here we refer to massive elliptical galaxies at the center of known clusters as BCGs, and others as NCEGs.}. We investigate whether the $P_{\rm cav}$--$M_{\rm mol}$ scaling relation is continuous from less massive NCEGs to giant elliptical galaxies (BCGs). If the relation is well established, it could be used in theoretical models of galaxy formation. For the $L_{1.4}$--$M_{\rm mol}$ and $L_{con}$--$M_{\rm mol}$ relations, we test for the existence of a correlation for the larger sample. We also investigate whether the mass of the black holes is a key factor in controlling AGN activity.

The paper is structured as follows:  Section~\ref{sec:reduc} describes the reduction of the ALMA data. Section~\ref{sec:result} presents the results of the data analysis and the correlation between the mass of molecular gas around black holes and AGN activity.  Section~\ref{sec:discuss} explores the implications of these correlations. We also investigate whether the black hole mass affects AGN activity, and discuss the source of the nonthermal emissions.  Section~\ref{sec:conc} is devoted to summarizing our findings. 

We assume $H_0=70\rm\: km\: s^{-1}\: Mpc^{-1}$, $\Omega_m=0.3$, and $\Omega_\Lambda=0.7$ in this study. All errors are $1\:\sigma$ unless otherwise noted.

\begin{deluxetable*}{cccccccccc}
\tablecaption{Target and observation details\label{tab:obs}}
\tablewidth{0pt}
\tablehead{\colhead{Target} &  \colhead{$z$} & \colhead{CO line} & \colhead{ALMA ID project} & \colhead{Obs. time} & \colhead{Date} & \colhead{Beam} & \colhead{PA} & \colhead{$v$ binning} & \colhead{rms} \\
\colhead{} & \colhead{} & \colhead{} & \colhead{} & \colhead{(min)} & \colhead{} & \colhead{($''$)} & \colhead{(deg)} & \colhead{($\rm km\: s^{-1}$)} & \colhead{($\rm mJy\: bm^{-1}$)} 
}
\startdata
NGC 4636 & 0.00313 & J=2-1 & 2015.1.00860.S & 59 & 2016-05-02 & 0.68 $\times$ 0.62 & -51 & 20 & 0.3 \\
NGC 4472 & 0.00327 & J=3-2 & 2017.1.00830.S & 17 & 2017-12-26 & 1.2 $\times$ 1.2 & 1 & 30 & 2 \\
NGC 4374 & 0.00339 & J=2-1 & 2013.1.00828.S & 51 & 2015-08-16 & 1.2 $\times$ 1.1 & -85 & 20 & 2 \\
NGC 5846 & 0.00571 & J=2-1 & 2015.1.00860.S & 40 & 2016-05-13 & 0.77 $\times$ 0.68 & -85 & 20 & 0.3 \\
NGC 1316 & 0.00601 & J=1-0 & 2019.1.01845.S & 71 & 2019-11-12 & 2.70 $\times$ 1.9 & -84 & 20 & 0.5 \\
NGC 5813 & 0.00653 & J=2-1 & 2015.1.00971.S & 67 & 2016-06-30 & 0.85 $\times$ 0.46 & -66 & 20 & 0.2 \\
NGC 4261 & 0.00726 & J=2-1 & 2017.1.00301.S & 31 & 2018-01-19 & 0.41 $\times$ 0.33 & 40 & 20 & 0.4 \\
NGC 7626 & 0.01136 & J=2-1 & 2019.1.00036.S & 9 & 2019-12-15 & 1.5 $\times$ 1.3 & 52 & 20 & 0.6 \\
IC 4296 & 0.01247 & J=2-1 & 2015.1.01572.S & 51 & 2016-06-04 & 0.61 $\times$ 0.57 & -80 & 20 & 0.4 \\
NGC 1600 & 0.01561 & J=3-2 & 2016.1.01135.S & 22 & 2016-12-27 & 0.77 $\times$ 0.48 & -73 & 20 & 0.7 \\
NGC 507 & 0.01646 & J=2-1 & 2016.1.00683.S & 42 & 2017-08-03 & 0.21 $\times$ 0.15 & 7 & 20 & 0.3 \\
NGC 315 & 0.01648 & J=2-1 & 2017.1.00301.S & 112 & 2018-10-19 & 0.34 $\times$ 0.19 & 9 & 20 & 0.3 \\
M 87 & 0.00428 & J=2-1 & 2013.1.00073.S & 45 & 2015-06-15 & 2.5 $\times$ 2.3 & -51 & 20 & 2 \\
\enddata
\end{deluxetable*}

\section{Sample Selection}
\label{sec:sample}

We examine CO line emission from cold molecular gas in elliptical galaxies using archival ALMA observations to investigate the correlation
between cold gas mass and AGN power. Our sample galaxies are taken from the
NCEGs studied by \citet{2010ApJ...720.1066C} and \citet{2011ApJ...735...11O}.
For 12 of them we found ALMA data that may contain CO line emissions \citep{2017ApJ...845..170B,2018ApJ...858...17T,2018ApJ...860....9M,2018MNRAS.475.3004S,2019MNRAS.484.4239R,2019PASJ...71...85M,2021ApJ...908...19B,2022A&A...664L..11S,2022ApJ...928..150T}. We also add M~87, which is a BCG not examined in Paper~I. The observational details of the final sample of 13 galaxies are given in Table~\ref{tab:obs}.  
For the NCEGs, their jet powers $P_{\rm cav}$ were obtained by \citet{2010ApJ...720.1066C}.  They defined the energy of each to be $4pV$, which is the enthalpy of a cavity filled with relativistic gas. The jet $P_{\rm cav}$ was obtained by dividing the energy by the age of the cavity (the buoyant rise time). If there are multiple cavities in a galaxy, their $P_{\rm cav}$ have been summed.
Since we have simply selected the targets that have been observed with ALMA and for which $P_{\rm cav}$ has been estimated, they are not a complete sample. 

The redshift range of our sample is 0.00313 (NGC~4636; Table~\ref{tab:obs}) to 0.1028 (PKS 0745-191; Paper~I) and we focus on the central 500~pc (1~kpc in diameter). In contrast, \citet{2019MNRAS.490.3025R} studied galaxies with a much wider range of redshifts ($z=0.0093$--0.596), and the beam sizes in which they estimated the circumnuclear gas mass varied among the galaxies (Figure~\ref{fig:rus19a}). The fixed radius (500~pc) and the narrower redshift range (Figure~\ref{fig:rus19a}) should make our sample less biased.

\section{Data Reduction}
\label{sec:reduc}

The sample galaxies were observed with ALMA at frequencies corresponding to the CO(J=1-0), CO(J=2-1), or CO(J=3-2)\footnote{We will refer to them as CO(1-0), CO(2-1), and CO(3-2).} rotational transition lines, mostly CO(J=2-1), with additional spectral windows that were used to image the millimeter/submillimeter (mm/submm; $\sim 100$--300 GHz) continuum emission.  The observations were single pointing, centered on the AGNs, except for NGC~1316, which was covered with mosaics. The data were calibrated using the appropriate version of the Common Astronomy Software Application (CASA) software \citep{2007ASPC..376..127M} and the ALMA Pipeline used for quality assurance. 

We subtracted the continuum emission using line-free channels with the CASA task \texttt{uvcontsub} (fitorder=1) to study the line emissions. We reconstructed the line cubes and the underlying continuum maps using the CASA task \texttt{tclean} with a threshold of $3\:\sigma$. We used Briggs weighting with a robust parameter of 2 (natural weighting), following previous studies \citep{2019A&A...631A..22O,2019MNRAS.490.3025R}. The synthesized beam size and rms in each final data cube are shown in Table~\ref{tab:obs}.

\begin{figure*}[ht!]
\plotone{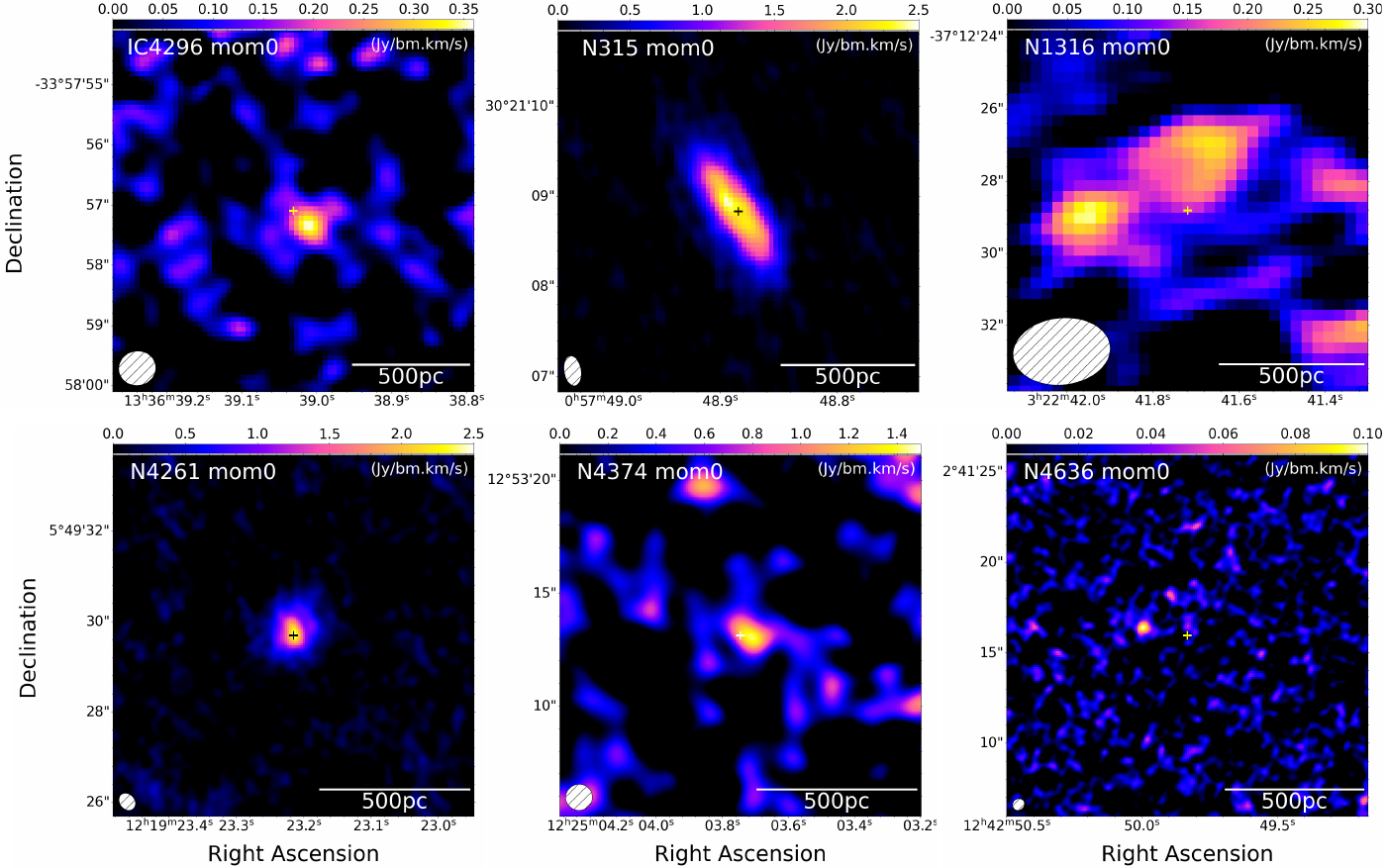}
\caption{Close-up CO integrated intensity (moment 0) map. 
The AGNs are located in the center of the figures and are indicated by the crosses. The beam is shown in white at the bottom left of each panel. }
\label{fig:image}
\end{figure*}

\section{Results}
\label{sec:result}

Using the CASA task \texttt{immoments}, we generate images of the integrated intensity for the AGN neighborhoods in the 13 galaxies that cover each CO line. From 6 of them (IC~4296, NGC~315, NGC~1316, NGC~4261, NGC~4374, and NGC~4636) we detected CO emission around their centers. Their images are shown in Figure~\ref{fig:image}.  
The following analysis focuses on these 6 galaxies. For the other 7 sources (6~NCEGs + M~87) we could not detect any cold gas, and only the upper limit of the mass is discussed.

Following Paper~I, we focus on the central 500~pc region because it is well resolved with ALMA and the gas in this region is likely to be fed into the black hole \citep{2008ApJ...681...73K,2022ApJ...924...24F}. For all 6 galaxies in Figure~\ref{fig:image}, the emission is concentrated on a scale smaller than $500$~pc. This is in contrast to the more massive BCGs studied in Paper~I, where the CO emission generally extends beyond the central 500~pc region.  
In the case of NGC~1316 and NGC~4636, the peak of the CO is offset from the location of the AGN (Figure~\ref{fig:image}), implying that the gas accretion is neither spherically nor axially symmetric. This happens when the accretion is chaotic \citep[Figure~13 in][]{2015A&A...579A..62G}. 

\begin{figure*}
\gridline{\fig{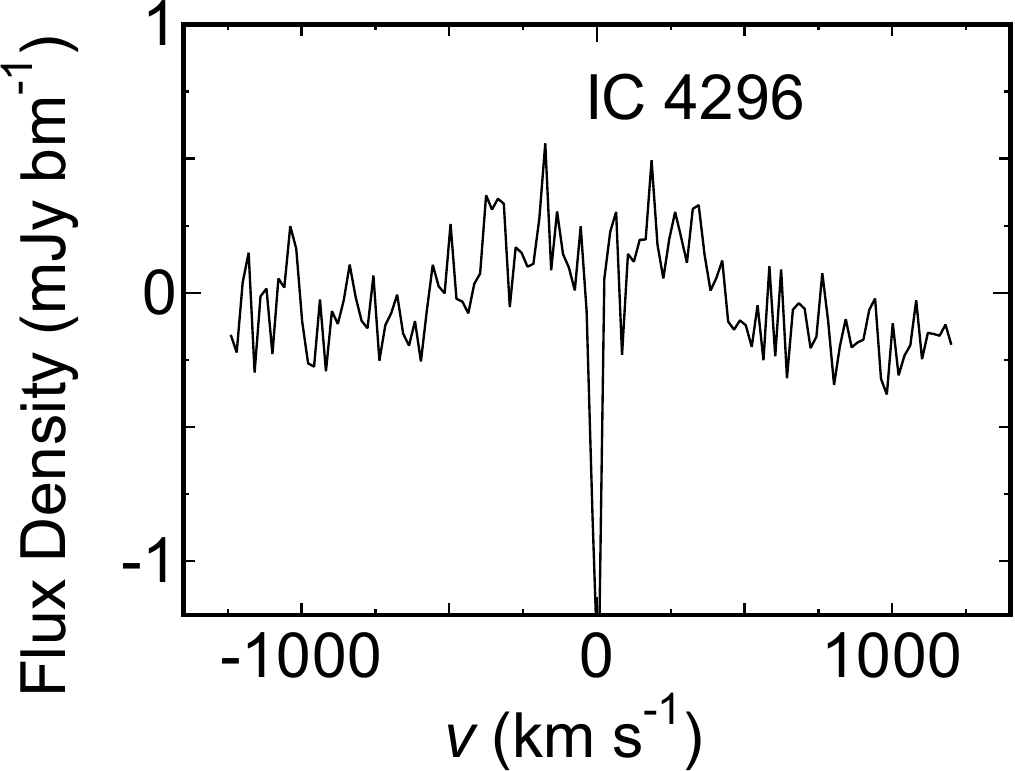}{0.3\textwidth}{}
          \fig{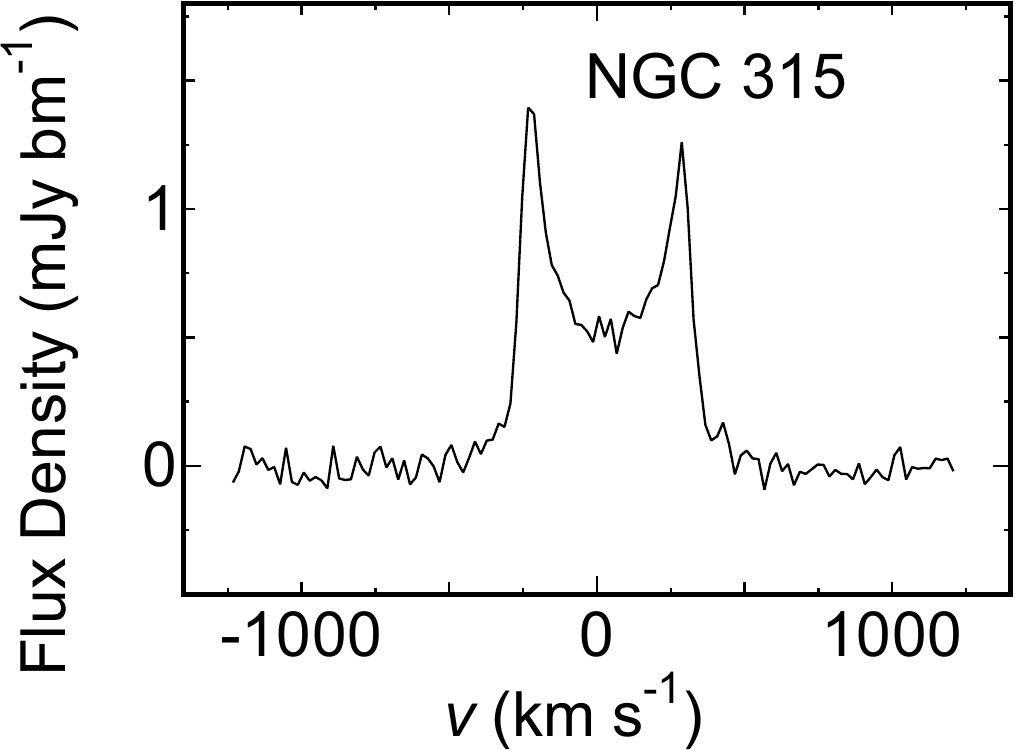}{0.3\textwidth}{}
          \fig{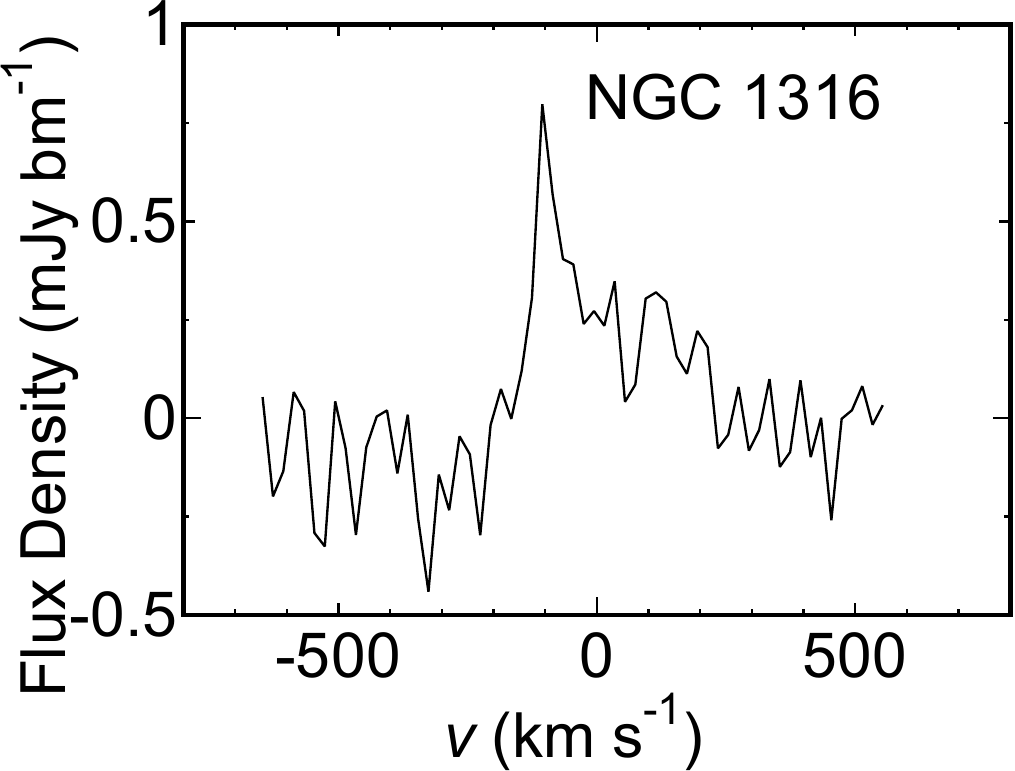}{0.3\textwidth}{}
          }
\gridline{\fig{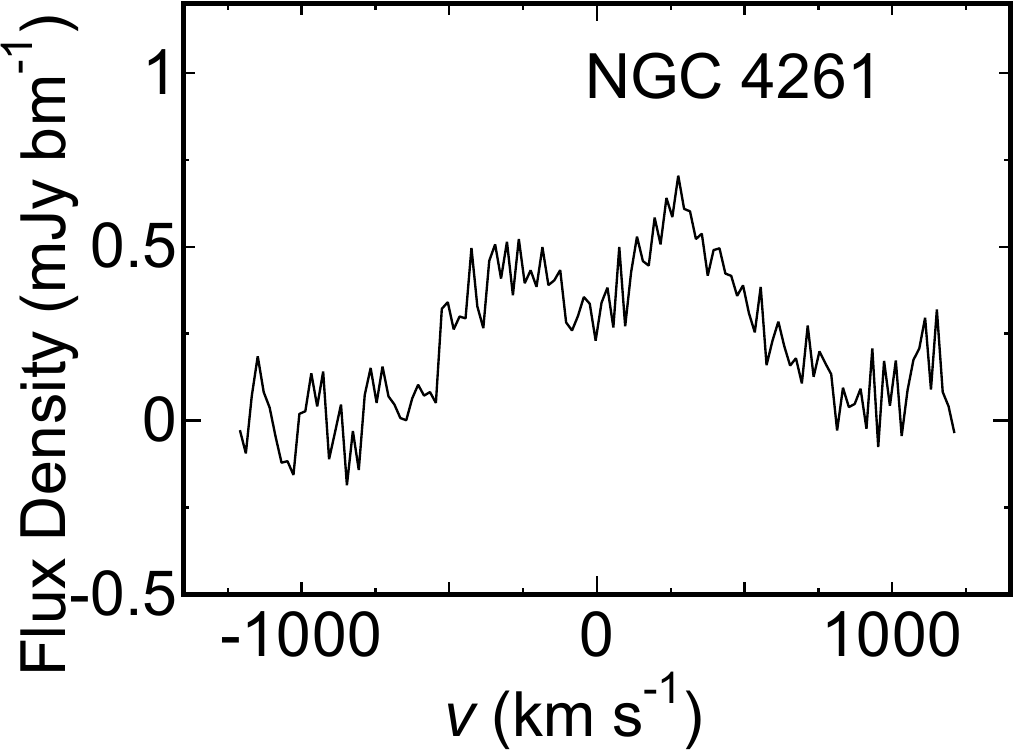}{0.3\textwidth}{}
          \fig{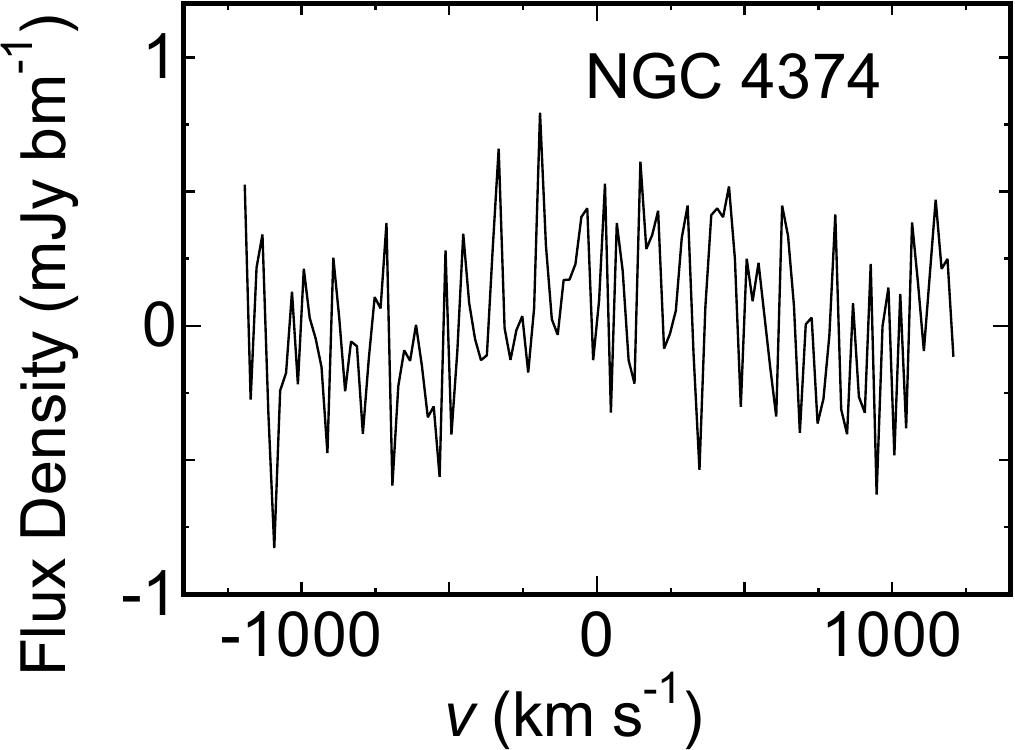}{0.3\textwidth}{}
          \fig{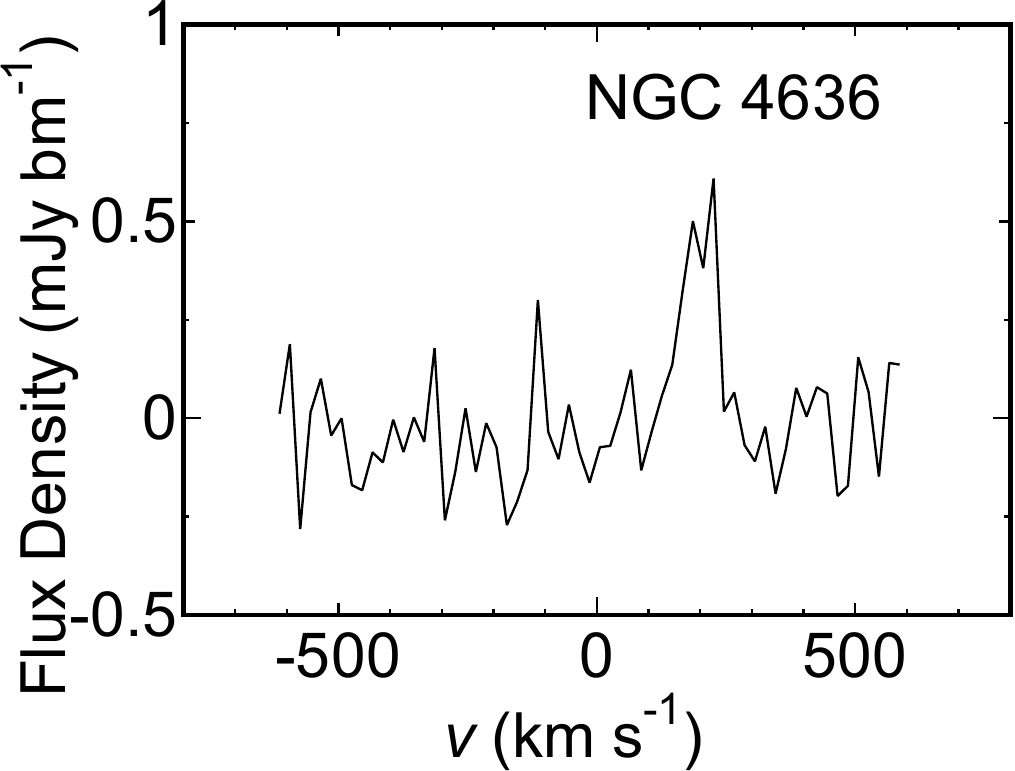}{0.3\textwidth}{}
          }
\vspace{-5mm}
\caption{CO line spectra within 500 pc from the AGNs.\label{fig:spec}}
\end{figure*}

We extract spectra from the emission regions for the 6 line-detected galaxies (Figure~\ref{fig:spec}) to check the profiles of the CO lines.
Unlike in Paper~I, we did not fit the spectra of the 6 galaxies with one or two Gaussian components, because not all of them are represented by Gaussians. For example, the spectrum of NGC~315 clearly shows the rotation of the gas and is not well fitted by Gaussians (Figure~\ref{fig:spec}).  
Instead, we derived the line flux densities from the brightness of the integrated intensity images (Figure~\ref{fig:image}). The images include channels where line emission is detected. For IC~4296 we ignore the channels corresponding to the prominent absorption (Figure~\ref{fig:spec}). We have confirmed that the two methods (Gaussian fit and intensity image) give consistent results when the spectrum can be well fitted by Gaussians.
For the 7 galaxies in which we could not detect cold gas, we estimate the $3\:\sigma$ upper limit of the line intensity by assuming that a possible line is within $\pm 500\:\rm km\: s^{-1}$ of the galaxy's velocity.  

In addition, unlike Paper~I, we do not discuss the mass accretion rate due to turbulent viscosity ($\dot{M}$), where the turbulent velocity is derived from the width of the Gaussian components. While $\dot{M}$ is determined by the molecular gas mass ($M_{\rm mol}$) and the turbulent velocity, as shown in Paper~I, the dependence of the turbulent velocity is not significant. This is because while $M_{\rm mol}$ varies by several orders of magnitude among galaxies, the turbulent velocity varies by only a factor of a few. As a result, $\dot{M}$ is approximately proportional to $M_{\rm mol}$. Moreover, the estimate of $\dot{M}$ has a large uncertainty, such as the gas morphology. Therefore, we focus on $M_{\rm mol}$ rather than $\dot{M}$ in this study.

\subsection{Molecular gas mass}
\label{sec:mass}

For galaxies in which CO emission is detected, we estimate the masses of the molecular gas from the CO intensities $S_{\rm CO}\Delta v$ given in Table~\ref{tab:obs2}. The table contains the objects studied in Paper~I. We take the following relation from \citet{2013ARA&A..51..207B}:
\begin{eqnarray}
 M_{\rm mol} &=& \frac{1.05\times 10^4}{F_{\rm ul}}\left(\frac{X_{\rm CO}}{2\times 10^{20}
\frac{\rm cm^{-2}}{\rm K\: km\: s^{-1}}}\right)
\left(\frac{1}{1+z}\right)\nonumber\\
& &\left(\frac{S_{\rm CO}\Delta v}{\rm Jy\: km\: s^{-1}}\right)
\left(\frac{D_{\rm L}}{\rm Mpc}\right)^2\: M_\odot\:,
\end{eqnarray}
where $X_{\rm CO}$ is the conversion factor from CO to H$_2$ and $D_{\rm L}$ is the luminosity distance. The empirical factor $F_{\rm ul}$ specifies the CO excitation used to estimate the CO(1-0) flux from higher J measurements, since we are comparing different transition lines. Specifically, $F_{21}$ gives a CO(2-1)/CO(1-0) line ratio on the flux density (Jy) scale of 3.2, and $F_{32}$ gives a CO(3-2)/CO(1-0) ratio of 7.2 \citep{2019MNRAS.485..229R,2019MNRAS.490.3025R}. Since the conversion factor $X_{\rm CO}$ is not well understood for elliptical galaxies, we adopt a default value of $X_{\rm CO}=2\times 10^{20}\rm\: cm^{-2}(K\: km\: s^{-1})^{-1}$ measured in the Milky Way, following previous studies (\citealt{2019A&A...631A..22O,2019MNRAS.490.3025R} and see their discussion). For galaxies in which no CO emission is detected, we estimate the upper bounds of the mass from the $3\:\sigma$ upper bounds of $S_{\rm CO}\Delta v$. The resulting mass $M_{\rm mol}$ is shown in Table~\ref{tab:obs2}.

\subsection{Circumnuclear gas and jet power}
\label{sec:Mmol-Pcav}

\begin{figure*}
\gridline{\fig{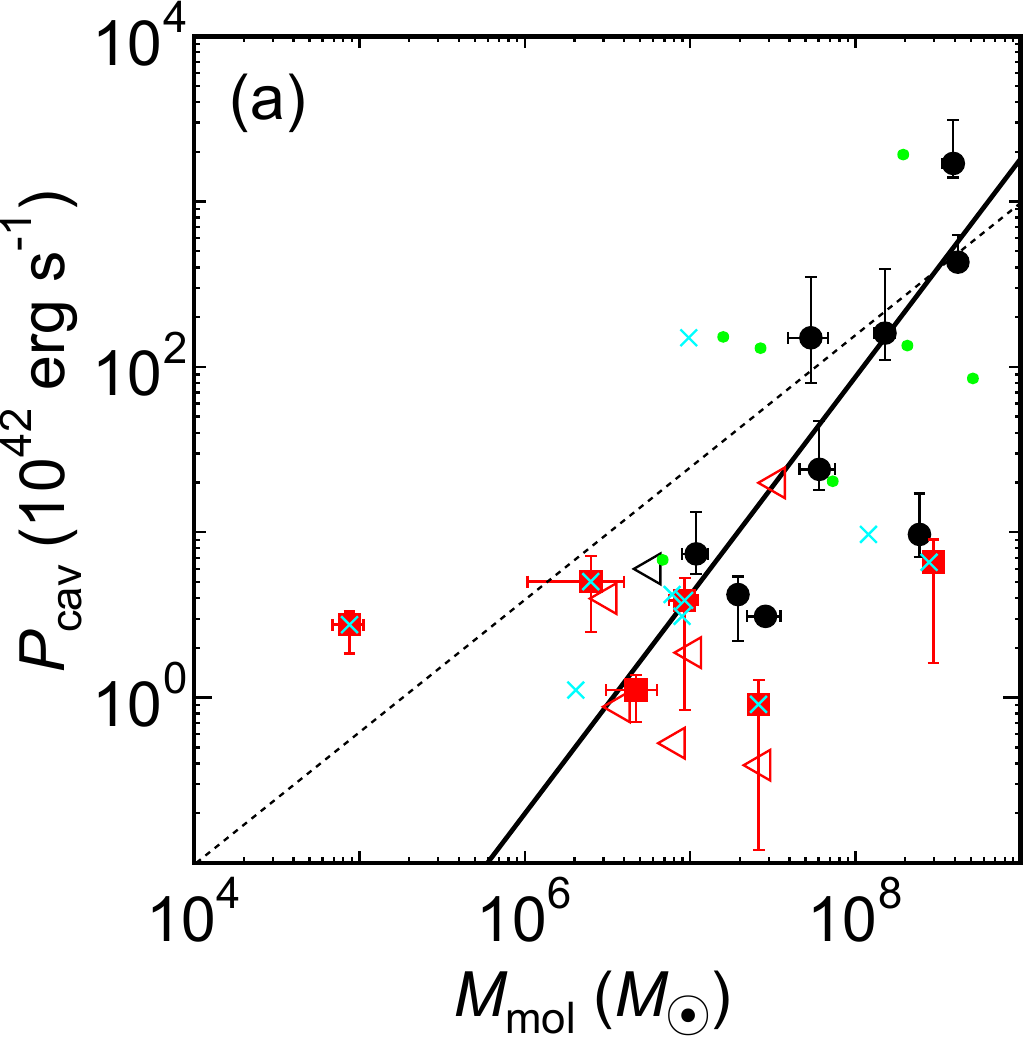}{0.3\textwidth}{}
          \fig{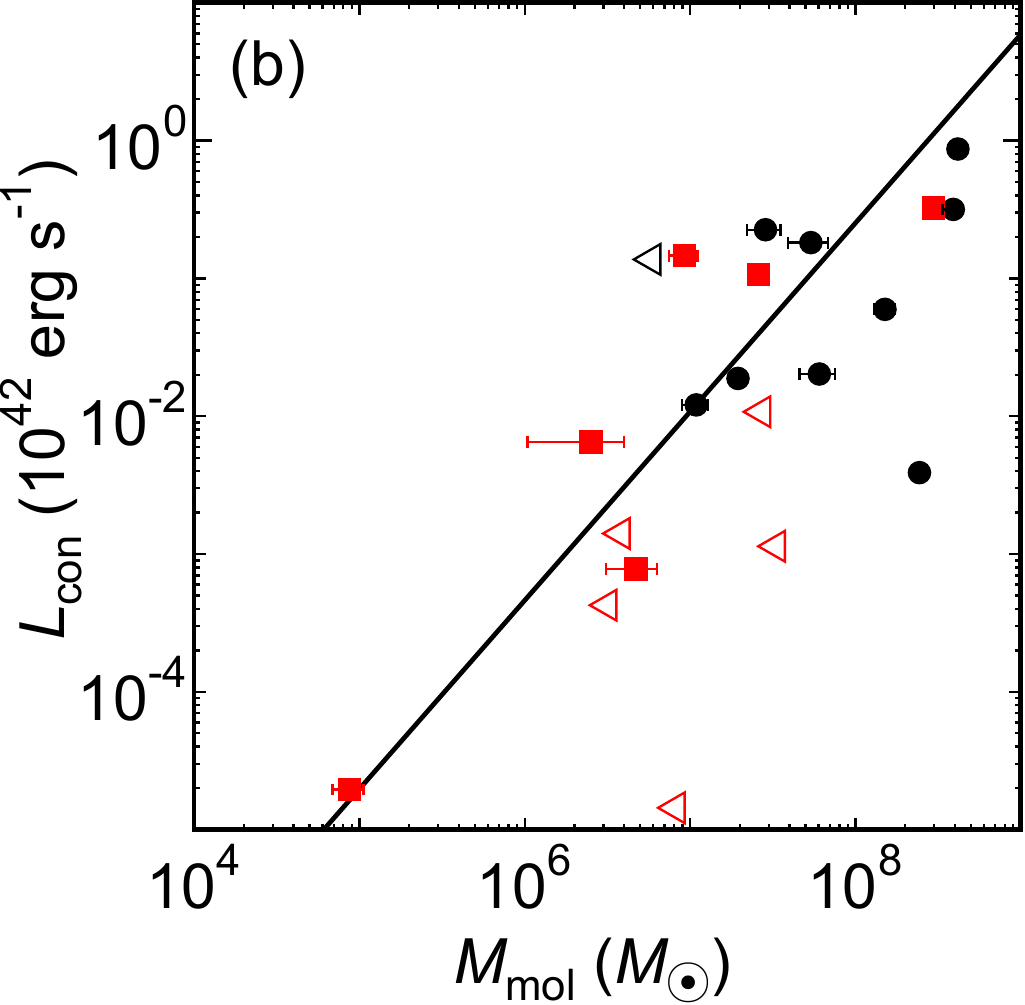}{0.3\textwidth}{}
          \fig{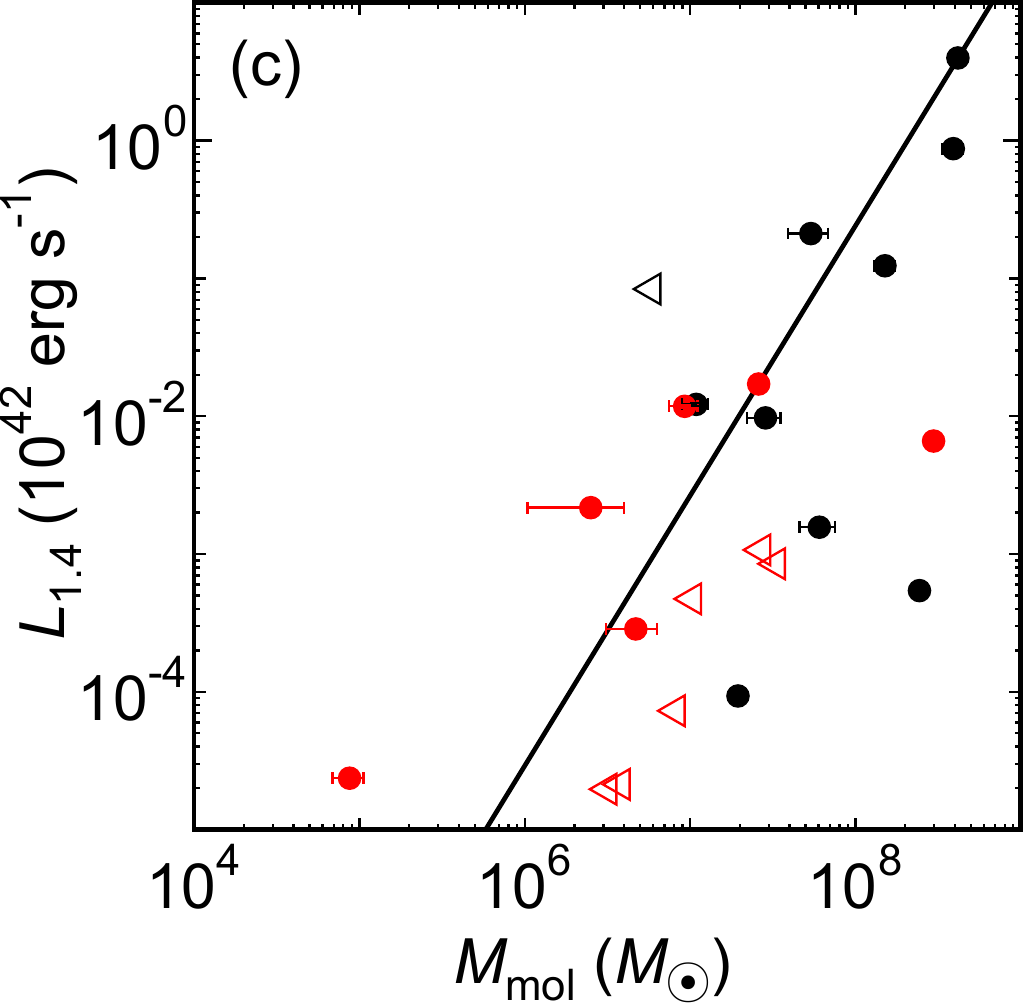}{0.3\textwidth}{}
          }
\vspace{-5mm}
\caption{Molecular gas mass within 500 pc of the AGN compared to (a) jet power estimated from X-ray cavities, (b) continuum luminosity in the mm/submm band, and (c) continuum luminosity at 1.4 GHz. 
Red filled squares represent NCEGs, while black filled circles represent BCGs.
Galaxies for which only the upper bounds of $M_{\rm mol}$ are determined are represented by the left-pointing open triangles (red denotes NCEGs and black denotes BCGs). The thick solid line shows the regression line (Akritas-Theil-Sen line). In (a), the cyan crosses are the objects for which molecular gas masses within 250~pc of the AGN are obtained, and the green dots are the values obtained by \citet{2019MNRAS.490.3025R}; uncertainties are not shown. For the former, objects for which only the upper limit is obtained are omitted. For the latter, the correlation found by \citet{2019MNRAS.490.3025R} is shown by the thin dotted line. \label{fig:Mmol}}
\end{figure*}

Figure~\ref{fig:Mmol}(a) shows the relationship between the mass of molecular gas within 500~pc of the center ($M_{\rm mol}$) and the AGN jet power required to produce all the observed X-ray cavities in the galaxy ($P_{\rm cav}$). Red filled squares represent NCEGs, while black filled circles represent BCGs. While the data of $P_{\rm cav}$ for the NCEGs are taken from \citet{2010ApJ...720.1066C}, those for the BCGs are taken from
\citet{2006ApJ...652...216R} and \citet{2018ApJ...853..177P}. The jet powers for the BCGs were measured in the same manner as for the NCEGs.

The NCEGs analyzed in this paper have lower values of $M_{\rm mol}$ ($\lesssim 10^7\: M_\sun$; Table~\ref{tab:obs2}) compared to the BCGs analyzed in Paper~I ($M_{\rm mol}\gtrsim 10^7\: M_\sun$). 
The Kendall rank correlation coefficient on the logarithmic scales is $\tau=0.472$ with a $p$-value of $1.9\times 10^{-3}$, which is below the standard threshold of 0.01. 
Here, the $p$ value indicates the probability that the quantities are statistically independent.
The coefficient and the regression line \citep[Akritas-Theil-Sen line;][]{NADAbook} are calculated for all galaxies including 7 galaxies for which only the upper limit of $M_{\rm col}$ was obtained using the R/CRAN package NADA\footnote{https://cran.r-project.org/web/packages/NADA/index.html}.
The regression line can be represented by a power law model as follows:
\begin{equation}
\label{eq:A1}
 \log\left(\frac{P_{\rm cav}}{10^{42}\rm erg\: s^{-1}}\right) = A_1\log\left(\frac{M_{\rm mol}}{M_\odot}\right) + B_1\:,
\end{equation}
where $A_1$ and $B_1$ are parameters, and we get $A_1 = 1.3$ and $B_1 = -8.6$. Thus, the relation can be written as $P_{\rm cav}\approx 4.1\times 10^{42}(M_{\rm mol}/10^7\: M_\sun)^{1.3}\rm\: erg\: s^{-1}$. The scatter in the relationship between $M_{\rm mol}$ and $P_{\rm cav}$ may be due to physical factors (see Section~\ref{sec:corr}). Note that the correlation is mainly influenced by the BCGs. This is because no correlation can be found for the 12 NCEGs alone.

We also checked the molecular gas mass within the radius of 250~pc. Considering the given beam sizes, we include all NCEGs and 5~BCGs (NGC~5044, Abell~262, Abell~3581, Abell~2052, and M~87). Hydra~A is not included due to strong absorption around the center (Paper~I). For the molecular gas mass within 250~pc (Figure~\ref{fig:Mmol}(a)), we examined the $P_{\rm cav}$--$M_{\rm mol}$ relation and found that the correlation coefficient is $\tau = 0.25$ and the $p$-value is $0.16$. The lack of correlation is partly due to the small number of BCGs that contribute much to the $P_{\rm cav}$--$M_{\rm mol}$ relationship.

\subsection{Circumnuclear gas and continuum emission}
\label{sec:cont}

While $P_{\rm cav}$ depends on AGN activity on a timescale of $\sim 10^7$~yr, the radio continuum emission is associated with recent ($\lesssim 10^3$~yr) AGN activity (Section~\ref{sec:intro}). Thus, the presence or absence of the correlation between $M_{\rm mol}$ and the continuum luminosity reflects the time variability of the gas supply and/or AGN activity on the shorter timescale.

Figure~\ref{fig:Mmol}(b) shows the relation between $M_{\rm mol}$ and $L_{\rm con}$, which is the continuum luminosity in the mm/submm band. The luminosity is calculated from the continuum flux in the line-free ALMA channels.  Fluxes observed at different frequencies are converted to those at the rest-frame CO(1-0) line frequency ($\nu_{10}=115.3$~GHz), assuming that the continuum emission is represented by a power law of $S_\nu \propto (\nu/\nu_{10})^\gamma$, where $\nu$ is the frequency. A typical spectral index of $\gamma=-0.75$ is assumed, following Paper~I. In fact, the contribution of dust emission is not significant (see Section~\ref{sec:origin}) and the results do not depend much on $\gamma$. For an object observed at the frequency $\nu_{\rm con}$ and located at the redshift $z$, we estimate the radio continuum luminosity at the frequency $\nu_{10}$ to be
\begin{equation}
 L_{\rm con} = \frac{4\pi D_{\rm L}^2 \nu_{\rm con} F_{\rm con}}{[(1+z)\eta]^{1+\gamma}}\:,
\end{equation}
where $F_{\rm con}$ is the observed flux (mJy), and $\eta$ is the line frequency ratios ($\eta=1$ for CO(1-0), $\eta\approx 2$ for CO(2-1), and $\eta\approx 3$ for CO(3-2)), which is needed to convert the flux between different line frequencies. The redshift dependence of $(1+z)^{-(1+\gamma)}$ comes from the $K$-correction \citep[e.g.][]{2022ApJ...924..133R}. The values of $\nu_{\rm con}$, $F_{\rm con}$, and $L_{\rm con}$ are listed in Table~\ref{tab:obs2}. 

The relationship between $L_{\rm con}$ and $M_{\rm mol}$ is shown in Figure~\ref{fig:Mmol}(b). There is a correlation between $M_{\rm mol}$ and $L_{\rm col}$ with a correlation coefficient of $\tau = 0.515$ and a $p$-value of $6.8\times 10^{-4}$. The regression line can be represented by a power law model of the form:
\begin{equation}
\label{eq:A2}
 \log\left(\frac{L_{\rm con}}{10^{42}\rm erg\: s^{-1}}\right) = A_2\log\left(\frac{M_{\rm mol}}{M_\odot}\right) + B_2\:,
\end{equation}
where $A_2$ and $B_2$ are fit parameters and we obtain $A_2=1.4$ and $B_2=-11.5$.

We also calculate the continuum luminosity at 1.4~GHz ($L_{\rm 1.4}$) from the observed flux at 1.4~GHz taken from the literature:
\begin{equation}
 L_{\rm 1.4} = \frac{4\pi D_{\rm L}^2 \nu_{\rm 1.4} F_{\rm 1.4}}
{(1+z)^{1+\gamma}}\:,
\end{equation}
where $\nu_{1.4}$ is $1.4$~GHz and $F_{1.4}$ is the observed flux at 1.4~GHz (Table~\ref{tab:obs2}), and $\gamma=-0.75$. The derived luminosities $L_{1.4}$ are shown in Table~\ref{tab:obs2} and compared to $M_{\rm mol}$ in Figure~\ref{fig:Mmol}(c). The correlation coefficient is $\tau = 0.437$ and the $p$-value is $4.0\times 10^{-3}$, indicating the existence of a correlation between $M_{\rm mol}$ and $L_{\rm 1.4}$, which can be represented by a power law model of the form:
\begin{equation}
\label{eq:A3}
 \log\left(\frac{L_{\rm 1.4}}{10^{42}\rm erg\: s^{-1}}\right) = A_3\log\left(\frac{M_{\rm mol}}{M_\odot}\right) + B_3\:,
\end{equation}
where $A_3$ and $B_3$ are fit parameters and we obtain $A_3=2.0$ and $B_3=-16.3$.

The presence of the correlations (equations~(\ref{eq:A2}) and~(\ref{eq:A3})) indicates that the short-term variability in gas supply and/or AGN activity is not strong enough to break the correlations.
In Paper~I we concluded that there were no correlations between $M_{\rm mol}$ and $L_{\rm col}$ and between $M_{\rm mol}$ and $L_{\rm 1.4}$\footnote{While the Pearson correlation coefficient was used in Paper~I, the Kendall rank correlation coefficient also rejects correlations.}. The results of the current paper suggest that the lack of correlations is simply due to the small number of sample galaxies and the narrow range of $M_{\rm mol}$.

\section{Discussion}
\label{sec:discuss}

\subsection{Correlations and Scatter}
\label{sec:corr}

The results shown in Section~\ref{sec:result} indicate that the mass of the circumnuclear molecular gas ($M_{\rm mol}$) is correlated not only with the jet power ($P_{\rm cav}$), but also with the continuum luminosities ($L_{\rm con}$ and $L_{\rm 1.4}$). 

The power $P_{\rm cav}$ depends on the averaged AGN activity on a timescale of $\sim 10^7$~yr, while the luminosities $L_{\rm con}$ and $L_{\rm 1.4}$ are associated with recent ($\lesssim 10^3$~yr) AGN activity (Section~\ref{sec:intro})\footnote{This does not necessarily mean that the luminosities are changing on a timescale of $\lesssim 10^3$~yr.}.  
The correlations shown in Figure~\ref{fig:Mmol} exhibit a scatter of about $\pm 0.5$~dex. This may indicate that the actual accretion of the cold gas is intermittent  due to turbulence in the gas \citep{2010MNRAS.408..961P,2017MNRAS.466..677G} or precipitation \citep{2015Natur.519..203V}. 

We note that the actual mass accretion rate ($\dot{M}$) or the AGN power depends not only on the mass of the circumnuclear gas ($M_{\rm mol}$), but also on other factors such as the dynamical stability of the gas around the black hole \citep{2008ApJ...681...73K,2022ApJ...924...24F}. The cycle of variation for each factor is likely to be different. 
For example, in the model of \citet{2022ApJ...924...24F}, the circumnuclear gas lifetime is $\gtrsim 10^8$~yr, while the timescale of dynamical stability is $\lesssim 10^7$~yr. Smaller timescales can also affect AGN activity. For example, if the circumnuclear gas is very clumpy on a scale much smaller than the resolution of ALMA, $\dot{M}$ could fluctuate on a timescale of $< 10^7$~yr. Indeed, recent observations by \citet{2023A&A...673A..52U} have shown that while the overall properties of the cool gas do not appear to change on timescales of $\gtrsim 10^7$ yr, AGN activity has shorter variability timescales.
If this difference changes $\dot{M}/M_{\rm mol}$ by a factor of $3~(\sim 10^{0.5})$, the scatter of the correlations in Figure~\ref{fig:Mmol} could be explained.  Moreover, the difference in the conversion factor $X_{\rm CO}$ among galaxies could be another cause of the scatter.

The difference in black hole spin among our sample galaxies may also contribute to the scatter in the relations in Figure~\ref{fig:Mmol}.
\citet{2007ApJ...658..815S} showed that elliptical galaxies are systematically more radio-loud than spiral galaxies, probably because of a larger black hole spin. 
The angular momentum brought by the accreted cold gas can spin up the black holes. In fact, some of our sample galaxies appear to have a disk (NGC~315 and NGC~4261 in Figures~\ref{fig:image} and~\ref{fig:spec}), showing that the gas has angular momentum. Other galaxies may also have unresolved small disks. It will be interesting to study the relationship between disk properties and AGN power in the future.

The relationship between the amount of molecular gas around the black hole and the activity of the AGN\footnote{Recently, \citet{2023arXiv231117848E} indicated that the circumnuclear molecular gas mass does not correlate with AGN activity. As they pointed out, the difference from our results could be explained by the difference in the samples. While our sample is limited to elliptical galaxies, their sample includes a wide range of galaxies and AGN activity.} is also known in disk galaxies \citep{2016ApJ...827...81I}. Therefore, the mechanism that fuels the AGN in elliptical galaxies may be similar to that in disk galaxies.

The NCEGs alone do not form a correlation in Figure~\ref{fig:Mmol}(a) (Section~\ref{sec:Mmol-Pcav}). Thus, it is not clear whether the $P_{\rm cav}$--$M_{\rm mol}$ relationship extends from the BCGs to the NCEGs. We need more samples to infer the correlation among the NCEGs. Although the number of the NCEG is small, the wide distribution of NCEGs in Figure~\ref{fig:Mmol}(a) suggests that the potential scatter of the possible correlation is large.

\begin{figure}[t]
\plotone{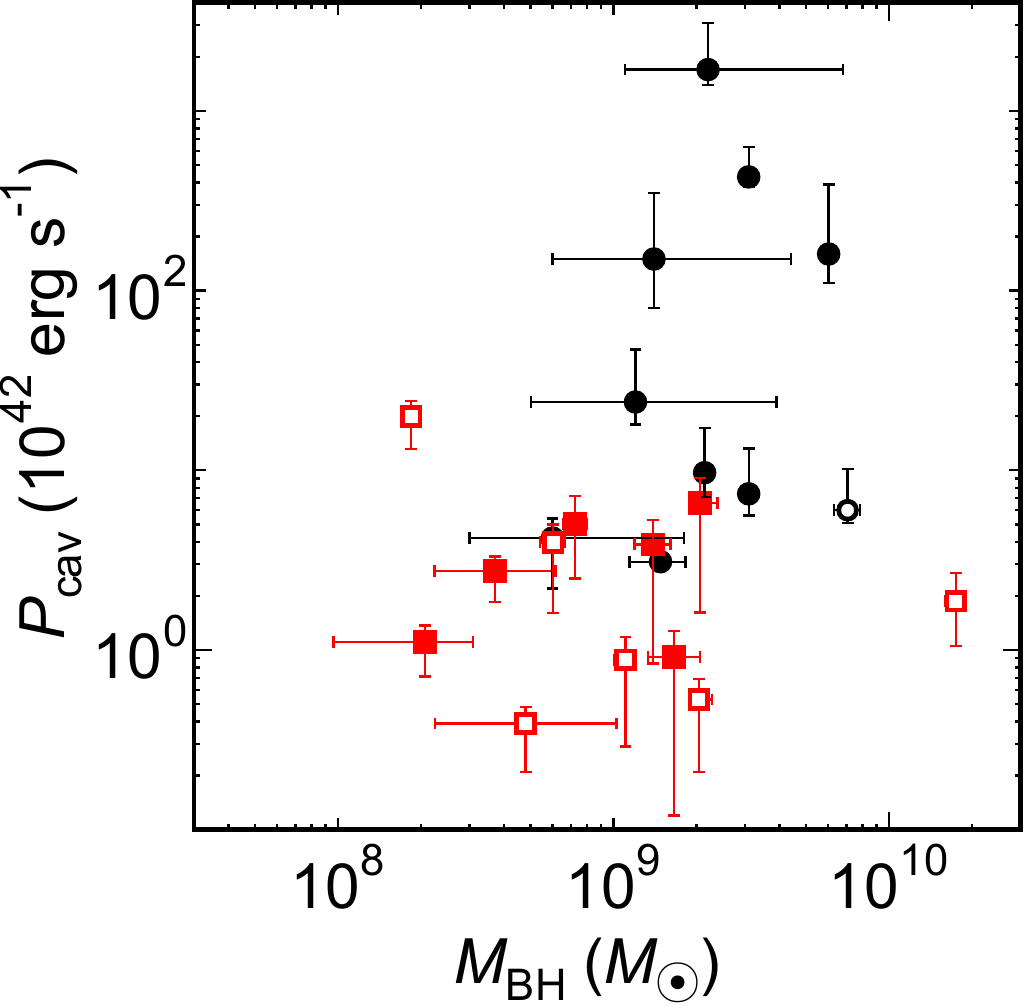} \caption{Jet power vs. black hole mass. Red squares
are the NCEGs and black circles are the
BCGs. Filled marks are the galaxies for which $M_{\rm mol}$ is
determined, while open marks are those for which only the upper limit of
$M_{\rm mol}$ is obtained.\label{fig:MBH-Pcav}}
\end{figure}

\begin{figure}[t]
\plotone{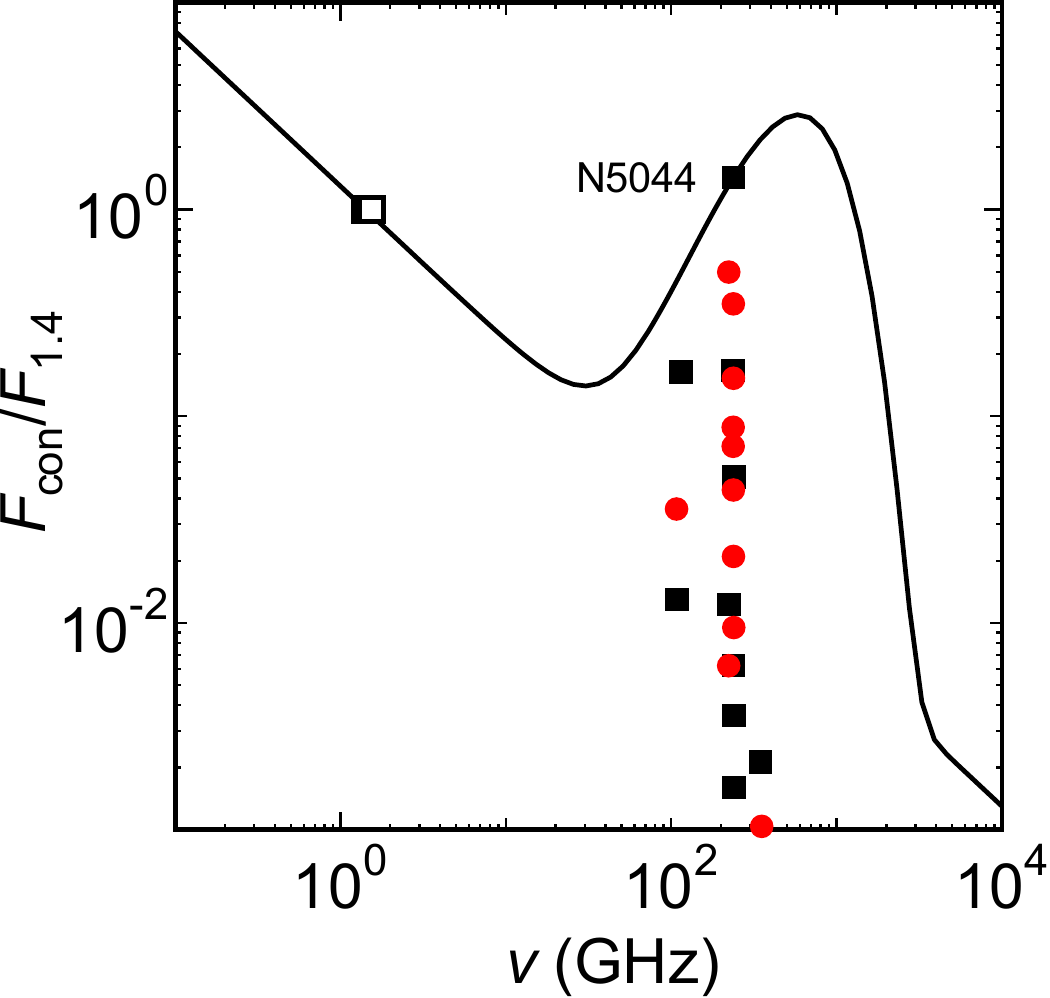} 
\caption{Spectra of continuum emission. Red filled squares (NCEGs) and black filled circles (BCGs) show $((1+z)\nu_{\rm con}, F_{\rm con}/F_{1.4})$, and open squares (NCEGs  + BCGs) show $((1+z)\nu_{1.4}, 1)$; the latter is almost independent of galaxies. The solid line represents a power-law + modified blackbody spectrum reproducing the observations for NGC~5044 (see text). The observational errors and the effects of $1+z$ are negligible.\label{fig:SED}}
\end{figure}

\subsection{Black hole mass}

Since the region we focused on (500~pc) is much smaller than the effective radii of the galaxies ($\gtrsim 10$~kpc; \citealt{2016JAsGe...5..277S}), it is unlikely that the overall properties of the galaxies directly affect $M_{\rm mol}$ and AGN activity. Instead, local properties such as the mass of the central black hole, $M_{\rm BH}$, may play some role in AGN activity. In fact, Figure~\ref{fig:Mmol}(a) shows that BCGs (black marks) tend to have larger $P_{\rm cav}$, and one might think that this is a result of their larger black hole masses. Here we check this possibility.

In Figure~\ref{fig:MBH-Pcav} we plot the jet power against the black hole masses $M_{\rm BH}$, which are taken from the literature and adjusted to our cosmological parameters. The masses are listed in Table~\ref{tab:obs2}. For all 22 galaxies, the correlation coefficient is $\tau = 0.234$, and the $p$-value is 0.13, indicating a lack of correlation. 
This is because some galaxies have significant uncertainties in $M_{\rm BH}$ and $P_{\rm cav}$ (Figure~\ref{fig:MBH-Pcav}).
Thus, there is no conclusive evidence that the mass of a black hole is a primary factor in the activity of an AGN, at least within our sample of galaxies. Of course, we do not deny a possible correlation between $M_{\rm BH}$ and $P_{\rm cav}$ for more precise data that would be obtained in the future. Since the mass of the black hole reflects the mass and velocity dispersion of the host galaxy \citep[e.g.][]{1998AJ....115.2285M,2000ApJ...539L...9F}, the fact that our sample galaxies have similar $M_{\rm BH}$ ($\sim 10^9\: M_\sun$) suggests that the host galaxies also have similar properties. 

For AGNs not restricted to those in elliptical galaxies, a correlation between jet power and black hole mass has been reported \citep{2006ApJ...637..669L}. However, the correlation has a large scatter ($\pm 1$~dex), while $M_{\rm BH}$ spans three orders of magnitude.  In addition, \citet{2015AJ....150....8C} pointed out that the correlation between jet power and black hole mass is unclear for AGNs with low gas accretion rates (small Eddington ratios), while the correlation between jet power and accretion rate is prominent for these AGNs. This may be consistent with our results because the accretion rates to AGNs in nearby elliptical galaxies are generally small \citep{2006ApJ...652...216R}.

\subsection{The origin of the mm/submm continuum emission}
\label{sec:origin}

We found the correlation between $M_{\rm mol}$ and continuum emission in section~\ref{sec:cont}. The origin of the continuum emission reflects what kind of matter is present around the AGNs.
In Figure~\ref{fig:SED} we show the ratio of $F_{\rm con}/F_{\rm 1.4}$. With the exception of NGC~5044, the ratio is less than one, indicating that most of the continuum emission is synchrotron emission from relativistic electrons.  In addition, we show a possible continuum spectrum (solid line), which is assumed to be represented by a combination of power-law (synchrotron) and modified blackbody (dust) components:
\begin{equation}
 S(\nu) = N_{\rm AGN}\nu^{\gamma} + N_{\rm BB}(1-e^{-(\nu/\nu_0)^b})B_\nu(\nu,T)\:,
\end{equation}
where $\gamma=-0.75$, $\nu_0=1.5$~THz, $b=1.5$, and $B_\nu(\nu,T)$ is a
blackbody distribution \citep[e.g.][]{2019A&A...621A..27F}. 
The dust temperature is assumed to be $T=10$~K (\citealt{2019ApJ...879..103F}; see Paper~I). The values for $N_{\rm AGN}$ and $N_{\rm BB}$ are adjusted to match the observations of NGC~5044. Figure~\ref{fig:SED} shows that, except for NGC~5044, the effect of the dust component is minimal at frequencies around 100--300 GHz, in agreement with local AGN observations \citep{2022ApJ...938...87K}.  A t-test shows that the ratios of $F_{\rm con}/F_{\rm 1.4}$ for the NCEGs are statistically indistinguishable from those for the BCGs.  In Paper I we discussed that galaxies with smaller $L_{\rm 1.4}$ may have a larger contribution from dust emission at $\sim 100$--300~GHz, because $L_{\rm con}/L_{\rm 1.4}\propto L_{\rm 1.4}^{-0.57}$. However, for the extended sample in this paper, the relationship is $L_{\rm con}/L_{\rm 1.4}\propto L_{\rm 1.4}^{-0.01}$, and the trend disappears.

\section{Conclusion}
\label{sec:conc}

Cold molecular gas has been observed in elliptical galaxies, and it is likely that this gas fuels their AGNs. To confirm this speculation, we measured the masses of the gas around the AGNs and discussed their relation to AGN activity. In this study, we extend the sample of galaxies studied in Paper~I to lower masses.

Using ALMA data, we studied the CO line emission within 500 pc of the center of 12 NCEGs plus one BCG, and detected the emission in 6 of them. We estimated the mass of the molecular gas $M_{\rm mol}$ from the emission and combined the results with those of 9 BCGs studied in Paper~I. We found that $M_{\rm mol}$ is correlated with the jet power of the central AGN ($P_{\rm cav}$), which is given by $P_{\rm cav}\approx 4.1\times 10^{42}(M_{\rm mol}/10^7\: M_\sun)^{1.3}\rm\: erg\: s^{-1}$,
although NCEGs alone do not show the correlation.
This suggests that AGNs are powered by molecular gas. Moreover, $M_{\rm mol}$ is also correlated with the continuum luminosity of AGNs at $\sim 1.4$~GHz and $\sim 100$--300~GHz. 
Since $P_{\rm cav}$ reflects long-term AGN activity, while the luminosity of the continuum indicates short-term AGN activity, our findings suggest that the amount of gas is a critical factor that affects AGN activity, regardless of the temporal scale.

On the other hand, we cannot find a clear relationship between the mass of black holes in AGNs ($M_{\rm BH}$) and $P_{\rm cav}$. This may be due to the large uncertainties in the data, but it suggests that $M_{\rm mol}$, rather than $M_{\rm BH}$, is the primary factor influencing AGN activity. We confirm that synchrotron radiation is mostly responsible for the continuum emission from AGNs at $\sim 1.4$--300 GHz. 

In galaxy formation studies, different models of gas accretion onto black holes have been adopted. For example, the Bondi accretion model is used in the Illustris simulations \citep[e.g.][]{2015MNRAS.452..575S}. Some semi-analytical models of galaxy formation use an empirical model of cold gas accretion \citep[e.g.][]{2014ApJ...794...69E}. The simple correlation we found between $M_{\rm mol}$ and $P_{\rm cav}$ could be used instead of these models.

\begin{acknowledgments}
We thank the reviewer and the statistician of AAS Journals for their useful comments.
 We also thank the EA ALMA Regional Center (EA-ARC) for their support. This work was supported by NAOJ ALMA Scientific Research Grant Code 2022-21A, and JSPS KAKENHI Grant Number JP22H01268, JP22H00158, JP22K03624 (Y.F.), JP20K14531, JP21H04496 (T.I.), JP19K03918 (N.Kawakatu), JP21H01137, JP18K03709 (H.N.), JP22K03686 (N.Kawanaka). This paper makes use of the following ALMA data: ADS/JAO.2011.0.00735.S, ADS/JAO.2012.1.00837.S, ADS/JAO.2013.1.00073.S, ADS/JAO.2013.1.00828.S, ADS/JAO2015.1.00598.S, ADS/JAO.2015.1.00623.S, ADS/JAO.2015.1.00627.S, ADS/JAO.2015.1.00644.S, ADS/JAO.2015.1.00860.S, ADS/JAO.2015.1.00971.S, ADS/JAO.2015.1.01198.S, ADS/JAO.2016.1.01214.S, ADS/JAO.2015.1.01572.S, ADS/JAO.2016.1.00683.S, ADS/JAO.2016.1.01135.S, ADS/JAO.2017.1.00301.S, ADS/JAO.2017.1.00830.S, ADS/JAO.2019.1.00036.S, ADS/JAO.2019.1.01845.S. ALMA is a partnership of ESO (representing its member states), NSF (USA) and NINS (Japan), together with NRC (Canada), MOST and ASIAA (Taiwan), and KASI (Republic of Korea), in cooperation with the Republic of Chile. The Joint ALMA Observatory is operated by ESO, AUI/NRAO and NAOJ. Data analysis was in part carried out on the Multi-wavelength Data Analysis System operated by the Astronomy Data Center (ADC), National Astronomical Observatory of Japan.
\end{acknowledgments}

\begin{longrotatetable}
\begin{deluxetable*}{cccccccccc}
\tablecaption{Target and observation details\label{tab:obs2}}
\tablewidth{0pt}
\tablehead{\colhead{Target} & \colhead{$S_{\rm CO}\Delta v$} & \colhead{$M_{\rm mol}$} & \colhead{$P_{\rm cav}$$^*$} & \colhead{$\nu_{\rm con}$} & \colhead{$F_{\rm con}$} & \colhead{$L_{\rm con}$} & \colhead{$F_{\rm 1.4}$$^\dagger$} & \colhead{$L_{\rm 1.4}$} & \colhead{$M_{\rm BH}$$^\ddagger$} \\
\colhead{} & \colhead{($\rm mJy bm^{-1} km s^{-1}$)} & \colhead{($10^7\: M_\odot$)} & \colhead{$(10^{42}\rm\: erg^{-1})$} & \colhead{(GHz)} & \colhead{(mJy)} & \colhead{$(10^{42}\rm\: erg^{-1})$} & \colhead{(mJy)} & \colhead{$(10^{42}\rm\: erg^{-1})$} & \colhead{($10^8\: M_\odot$)} 
}
\startdata
NGC 4636 & $140\pm 31$ & $0.0087\pm 0.0018$ & $2.76_{-0.91}^{+0.56}$$^a$ & 221.8 & $0.483\pm 0.032$ & $(2.0\pm 0.1)\times 10^{-5}$ & $77.8\pm 2.8$$^d$ & $(2.4\pm 0.1)\times 10^{-5}$ & $3.7_{-1.5}^{+2.4}$$^h$ \\
NGC 4472 & $<12700$ & $<0.82$ & $0.53_{-0.32}^{+0.16}$$^a$ & 351.2 & $0.228\pm 0.039$ & $(1.4\pm 0.2)\times 10^{-5}$ & $220\pm 8$$^d$ & $(7.3\pm 0.3)\times 10^{-5}$ & $20_{-1}^{+2}$$^i$ \\
NGC 4374 & $3620\pm 590$ & $0.25\pm 0.15$ & $5.03_{-2.53}^{+2.17}$$^a$ & 236.9 & $127.9\pm 5.2$ & $(6.5\pm 0.3)\times 10^{-3}$ & $6100\pm 150$$^e$ & $(2.2\pm 0.1)\times 10^{-3}$ & $7.2_{-0.7}^{+0.8} $$^j$ \\
NGC 5846 & $<1930$ & $<0.38$ & $0.88_{-0.59}^{+0.30}$$^a$ & 221.3 & $10.46\pm 0.31$ & $(1.4\pm 0.0)\times 10^{-3}$ & $21\pm 1$$^d$ & $(2.1\pm 0.1)\times 10^{-5}$ & $11_{-1}^{+1}$$^k$ \\
NGC 1316 & $673\pm 230$ & $0.47\pm 0.16$ & $1.11_{-0.40}^{+0.26}$$^a$ & 106.9 & $9.06\pm 0.15$ & $(7.8\pm 0.1)\times 10^{-4}$ & $25\pm 10$$^d$ & $(2.9\pm 0.1)\times 10^{-4}$ & $2.1_{-1.1}^{+1.0}$$^k$ \\
NGC 5813 & $<1230$ & $<0.32$ & $3.97_{-2.36}^{+1.02}$$^a$ & 236 & $2.26\pm 0.11$ & $(4.3\pm 0.2)\times 10^{-4}$ & $14.8\pm 1.0$$^d$ & $(2.0\pm 0.1)\times 10^{-5}$ & $6.0_{-0.6}^{+0.6}$$^k$ \\
NGC 4261 & $8140\pm 290$ & $2.6\pm 0.1$ & $0.91_{-0.79}^{+0.37}$$^a$ & 235.6 & $459\pm 48$ & $0.11\pm 0.01$ & $10400\pm 200$$^e$ & $(1.7\pm 0.0)\times 10^{-2}$ & $17_{-3}^{+4}$$^l$ \\
NGC 7626 & $<842$ & $<2.7$ & $0.39_{-0.18}^{+0.09}$$^a$ & 234.1 & $18.96\pm 0.58$ & $(1.1\times 0.0)\times 10^{-2}$ & $265\pm 8$$^d$ & $(1.1\pm 0.0)\times 10^{-3}$ & $4.8_{-2.6}^{+5.5}$$^h$ \\
IC 4296 & $630\pm 195$ & $0.93\pm 0.18$ & $3.87_{-3.03}^{+1.44}$$^a$ & 233.8 & $214\pm 12$ & $0.15\pm 0.01$ & $2420\pm 50$$^e$ & $(1.2\pm 0.0)\times 10^{-2}$ & $14_{-2}^{+2}$$^m$ \\
NGC 1600 & $<1560$ & $<1.0$ & $1.87_{-0.82}^{+0.82}$$^a$ & 340.5 & $< 0.270$ & $<3.8\times 10^{-4}$ & $61.6\pm 2.6$$^d$ & $(4.7\pm 0.2)\times 10^{-4}$ & $170_{-20}^{+20}$$^n$ \\
NGC 507 & $<2010$ & $<3.3$ & $19.9_{-6.8}^{+4.4}$$^a$ & 234.9 & $0.945\pm 0.07$ & $(1.1\pm 0.1)\times 10^{-3}$ & $99.5\pm 1.5$$^f$ & $(8.5\pm 0.1)\times 10^{-4}$ & $1.8_{-0.0}^{+0.0}$$^o$ \\
NGC 315 & $18000\pm 445$ & $30\pm 1$ & $6.58_{-4.96}^{+2.48}$$^a$ & 233.6 & $270\pm 13$ & $0.32\pm 0.02$ & $772\pm 25$$^d$ & $(6.6\pm 0.2)\times 10^{-3}$ & $21_{-1}^{+3}$$^l$ \\ \hline
M 87 & $<5290$ & $<0.59$ & $6.0_{-0.9}^{+4.2}$$^b$$^e$ & 222.8 & $1807\pm 78 $ & $0.14\pm 0.01$ & $147000\pm 5000$$^e$ & $(8.4\pm 0.3)\times 10^{-2}$ & $71_{-8}^{+8}$$^p$ \\
NGC 5044 & $517\pm 71$ & $2.0 \pm 0.2$ & $4.2_{-2.0}^{+1.2}$$^c$ & 235 .2 & $49.5 \pm 3.7$ & $(1.9\pm 0.1)\times 10^{-2}$ & $34.7\pm 1.1$$^d$ & $(9.3\pm 0.3)\times 10^{-5}$ & $6.0_{-3.0}^{+12.0}$$^q$ \\
Centaurus & $208 \pm 37$ & $1.1 \pm 0.2$ & $7.4_{-1.8}^{+5.8}$$^b$ & 107 .1 & $51.6 \pm 2.6$ & $(1.2\pm 0.1)^{-2}$ & $3980\pm 110$$^g$ & $(1.2\pm 0.0)\times 10^{-2}$ & $31_{-1}^{+1}$$^r$ \\
Abell  262 & $1420\pm 100$ & $24 \pm 2$ & $9.7_{-2.6}^{+7.5}$$^c$ & 235 .6 & $3.336 \pm 0.069$ & $(3.9\pm 0.1)\times 10^{-3}$ & $65.7\pm 2.3$$^d$ & $(5.4\pm 0.2)\times 10^{-4}$ & $21_{-1}^{+1}$$^r$ \\
Abell 3581 & $111\pm 25$ & $2.9 \pm 0.7$ & $3.1$$^c$ & 231 .8 & $107.5 \pm 3.7$ & $0.23\pm 0.01$ & $646\pm 23$$^d$ & $(9.7\pm 0.3)\times 10^{-3}$ & $15_{-3}^{+3}$$^r$ \\
Abell  2052 & $43 \pm 11$ & $5.4 \pm 1.5$ & $150_{-70}^{+200}$$^c$ & 229 .5 & $34.3 \pm 3.1$ & $0.18\pm 0.02$ & $5500\pm 210$$^d$ & $(2.1\pm 0.1)\times 10^{-1}$ & $14_{-8}^{+30}$$^q$ \\
2A0335+096 & $262 \pm 63$ & $6.1 \pm 1.5$ & $24_{-6}^{+23}$$^b$ & 110 .5 & $6.01 \pm 0.34$ & $(2.0\pm 0.1)\times 10^{-2}$ & $36.7\pm 1.8$$^d$ & $(1.6\pm 0.1)\times 10^{-3}$ & $12_{-7}^{+27}$$^q$ \\
Hydra A & $460\pm 26$ & $42 \pm 2$ & $430_{-50}^{+200}$$^c$ & 227 .6 & $65.23 \pm 0.19$ & $0.87\pm 0.0$ & $40800\pm 1300$$^d$ & $4.0\pm 0.1$ & $31_{-1}^{+1}$$^r$ \\
Abell  1795 & $654\pm 93$ & $15 \pm 2$ & $160_{-50}^{+230}$$^b$ & 225 .8 & $3.29 \pm 0.041$ & $(6.0\pm 0.1)\times 10^{-2}$ & $925\pm 28$$^d$ & $(1.2\pm 0.0)\times 10^{-1}$ & $60_{-4}^{+4}$$^r$ \\
PKS 0745-191 & $353 \pm 49$ & $39 \pm 5$ & $1700_{-300}^{+1400}$$^c$ & 314 .6 & $5.03 \pm 0.35$ & $0.32\pm 0.02$ & $2370\pm 80$$^d$ & $(8.6\pm 0.3)\times 10^{-1}$ & $22_{-11}^{+46}$$^q$ \\
\enddata
\tablecomments{Galaxies from NGC~4636 to NGC~315 are NCEGs and those from M~87 to PKS 0745-191 are BCGs.}
$^*$References of $P_{\rm cav}$: ($a$) \citet{2010ApJ...720.1066C}, ($b$) \citet{2006ApJ...652...216R}, ($c$) \citet{2018ApJ...853..177P}.\\
$^\dagger$References of $F_{\rm 1.4}$: ($d$) \citet{1998AJ....115.1693C}, ($e$) \citet{2014MNRAS.440..696A}, ($f$) \citet{2011AandA...526A.148M}, ($g$) \citet{1981AandAS...45..367K}.\\
$^\ddagger$References of $M_{\rm BH}$: ($h$) \citet{2016ApJ...831..134V}, ($i$) \citet{2013AJ....146...45R}, ($j$) \citet{2010ApJ...721..762W}, ($k$) \citet{2013ApJ...764..151G}, ($l$) \citet{2021ApJ...908...19B}, ($m$) \citet{2009ApJ...690..537D}, ($n$) \citet{2016Natur.532..340T}, ($o$) \citet{2019ApJ...875..141P},
\end{deluxetable*}
\vspace{-6.5mm}
\hspace{-72mm}
\footnotesize{($p$) \citet{2019ApJ...875L...1E}, ($q$) \citet{2017MNRAS.464.4360M}}, ($r$) \citet{2018MNRAS.474.1342M}.
\end{longrotatetable}

\bibliography{mmol}{} 
\bibliographystyle{aasjournal}

\end{document}